\documentclass[preprint,eqsecnum,floats,aps,showpacs,superscriptaddress]{revtex4}
\def\thalf{{\textstyle{3\over 2}}}
\def\ohalf{{\textstyle{1\over 2}}}
\input{epsf}
\usepackage{epsfig}
\usepackage{amsfonts}
\usepackage{amssymb}
\begin{document}

\title{Extraction and Interpretation of $\gamma N \rightarrow \Delta$ Form Factors \\ 
 within a Dynamical Model  \\

(From EBAC, Thomas Jefferson National Accelerator Facility)}

\author{B. Juli\'a-D\'{\i}az}
\affiliation{Excited Baryon Analysis Center, Thomas Jefferson National Accelerator
Facility, Newport News, Va. 22901}
\affiliation{Departament d'Estructura i Constituents de la Mat\`{e}ria,
Universitat de Barcelona, E--08028 Barcelona, Spain}

\author{T.-S. H. Lee}
\affiliation{Excited Baryon Analysis Center, Thomas Jefferson National Accelerator
Facility, Newport News, Va. 22901}
\affiliation{Physics Division, Argonne National Laboratory, 
Argonne, IL 60439}

\author{T. Sato}
\affiliation{Excited Baryon Analysis Center, Thomas Jefferson National Accelerator
Facility, Newport News, Va. 22901}
\affiliation{Department of Physics, Osaka University, Toyonaka, 
Osaka 560-0043, Japan}

\author{L.C. Smith}
\affiliation{Excited Baryon Analysis Center, Thomas Jefferson National Accelerator
Facility, Newport News, Va. 22901}

\affiliation{Physics Department, University of Virginia, Charlottesville, Va. 22901}

\begin{abstract}
Within the dynamical model of Refs. [Phys. Rev. C54, 2660 (1996); C63, 055201 (2001)],
we perform an analysis of recent data of pion electroproduction 
reactions at energies near the $\Delta(1232)$ resonance.
We discuss possible interpretations of the extracted bare and 
dressed $\gamma N \rightarrow \Delta$ form factors in terms of
relativistic constituent quark models and Lattice QCD calculations.
Possible future developments are discussed.
\end{abstract}

\thanks{Notice: Authored by Jefferson Science Associates, LLC under U.S. DOE
Contract No. DE-AC05-06OR23177. The U.S. Government retains a
non-exclusive, paid-up, irrevocable, world-wide license to publish or
reproduce this manuscript for U.S. Government purposes.}
\renewcommand{\thefootnote}{\arabic{footnote}}

\maketitle

\section{Introduction}

One of the main challenges in hadron physics is to understand
within Quantum Chromodynamics (QCD) the structure of the nucleon 
and its excited states. Through their couplings with the meson-baryon 
continuum, these excited states may be identified with 
the nucleon resonances ($N^*$) in various meson-baryon reactions. 
Ideally, one would like to study the $N^*$ structure by analyzing 
the meson-baryon reaction data completely within QCD. This 
however is far from our reach. As a compromise, one can analyze 
meson-baryon reaction data by developing reaction models within 
which both the internal structure of hadrons and the reaction mechanisms 
are modeled using theoretical guidances deduced from our current 
understanding of QCD. Of course, the separation between the 
reaction mechanisms and the internal structure of hadrons is 
by no means well defined theoretically. Thus the model dependence 
of the interpretations of the extracted $N^*$ parameters is 
unavoidable. This is not very satisfactory, but seems to be the 
only option we have at the present time. With this understanding, 
many models have been developed in recent years to analyze the very 
extensive electromagnetic meson production data which have been 
obtained by intense experimental efforts at Jefferson Laboratory (JLab), MIT-Bates,
LEGS, Mainz, GRAAL, Bonn, and SPring-8. For a comprehensive review 
of the present experimental and theoretical status we refer the reader 
to Ref.~\cite{burkertlee04}.  

In this work we focus on recent measurements of pion electroproduction at energies
near the $\Delta(1232)$ resonance, and in particular the data at $Q^2<$~2.00~(GeV/c)$^2$
obtained at JLab~\cite{joo02,smi06}, MIT-Bates~\cite{bates} and 
Mainz~\cite{leu01,Sta06,Spa06}.  One possible approach to analyze these data is to apply
the dynamical model developed by Sato and Lee (SL) in Refs.~\cite{satolee96,satolee01}. 
The SL model, as well as the unitary isobar model MAID~\cite{maid}, and the 
Dubna-Mainz-Taipei (DMT) model~\cite{dmt} 
have been used extensively to extract the $\gamma N \rightarrow \Delta$ form factors from the
data. The SL model gives reasonably good descriptions of the very extensive
data accumulated since 2001, using parameters obtained by analyzing the pion 
photoproduction data from LEGS~\cite{legs} and  Mainz~\cite{mainz}, and the pion
electroproduction data at $Q^2=2.8, 4.0$ (GeV/c)$^2$ from JLab~\cite{fro99}. 
In this paper, we would like to re-visit this model and to include the 
recent data~\cite{joo02,smi06,bates,leu01,Sta06,Spa06}  in
a comprehensive analysis to extract more precisely the $\gamma N\rightarrow\Delta$
form factors.  Furthermore, we explore the theoretical interpretations of
the extracted $\gamma N\rightarrow\Delta$ form factors 
in terms of two relativistic
quark model calculations~\cite{capstick-1,bruno-1,bruno-2} and a recent Lattice QCD
calculation~\cite{alexandrou}.

In section II, we recall the essential ingredients of the
SL model and present a refinement of the model from 
improvement of fits to the $\pi N$ scattering data.
 Section III is devoted to defining explicitly
the $\gamma N \rightarrow \Delta$ transition form factors which are the 
focus of this work.
The extraction of $\gamma N\rightarrow \Delta$ form factors from
electroproduction data is discussed in section IV.
We then discuss possible theoretical interpretations of the extracted
$\gamma N\rightarrow \Delta$ form factors in section V.
In section VI, we give a summary and discuss possible future 
developments.

\section{The SL model and $\pi N$ scattering} 
The essential feature of 
the SL model is to have a consistent description of both the $\pi N$ 
scattering and the electromagnetic pion production reactions. 
This is achieved by 
applying a unitary transformation method~\cite{kso,satolee96}
 to derive an effective Hamiltonian 
from the interaction Lagrangians with $N$, $\Delta$, $\pi$,
$\rho$ ,$\omega$, and photon fields. The details of the model can be seen in 
Ref.~\cite{satolee96,satolee01}. 
For our present purposes, it is sufficient to just recall the expression
of the $\gamma N \rightarrow \pi N$ amplitude $T_{\pi N,\gamma N}$
given in Ref.~\cite{satolee96}
\begin{eqnarray}
T_{\pi N,\gamma N}(E) &=& t_{\pi N,\gamma N}(E) + t^R_{\pi N,\gamma N}(E)\,,
\label{eq:tpingn}
\end{eqnarray}
where the non-resonant amplitude is defined by
\begin{eqnarray}
\label{eq:nr-pipi}
t_{\pi N,\gamma N}(E)=[t_{\pi N,\pi N}(E)G_{\pi N}(E)+1]v_{\pi N,\gamma N}
\label{eq:nr-tpingn}\,
\end{eqnarray}
with the non-resonant $\pi N$ scattering amplitude defined by
\begin{eqnarray}
t_{\pi N,\pi N}(E)=v_{\pi N,\pi N}[1
+ G_{\pi N}(E)t_{\pi N,\pi N}(E)] \,.
\label{eq:nr-tpinpin}
\end{eqnarray} 
Here  $v_{\pi N,\pi N}$ is a $\pi N$ potential
and $G_{\pi N}(E)$ is a $\pi N$ propagator with
 relativistic kinetic energy operators.

The resonant amplitude is given by
\begin{eqnarray}
t^R_{\pi N,\gamma N}(E)=
\frac{\bar{\Gamma}^\dagger_{\Delta,\pi N}
\bar{\Gamma}_{\Delta,\gamma N}}
{E- m_\Delta -\Sigma_\Delta(E)}  \,,
\label{eq:rgpi}
\end{eqnarray}
where the dressed vertex interactions are defined, to the
first order in electromagnetic coupling, by
\begin{eqnarray}
\bar{\Gamma}^\dagger_{\Delta,\pi N}&=& [t_{\pi N,\pi N}(E)G_{\pi N}(E) + 1]
{\Gamma^\dagger}_{\Delta,\pi N} \,,
\label{eq:f-pind}\\
\bar{\Gamma}_{\Delta,\gamma N}&=& \Gamma_{\Delta,\gamma N}+
{\Gamma}_{\Delta,\pi N} G_{\pi N}(E) t_{\pi N,\gamma N} \,.
\label{eq:f-dgn} 
\end{eqnarray}
In the above equations,
 $\Gamma_{\pi N, \Delta}$ and $\Gamma_{\Delta,\gamma N}$ are the
bare vertex interactions.
The energy shift $\Sigma_\Delta(E)$ is calculated from
\begin{eqnarray}
\Sigma_\Delta(E) = \Gamma_{\Delta,\pi N}G_{\pi N}
\bar{\Gamma}^\dagger_{\Delta,\pi N} \, .
\label{eq:dself}
\end{eqnarray}
Here we remark that the $\Delta$ propagator of 
Eq.~(\ref{eq:rgpi}) is the mathematical consequence of the Hamiltonian formulation of
the reactions. It differs from the covariant forms, mainly used in tree-diagram models 
and extensively discussed in the literature on possible off-shell 
effects~\cite{mukhopadhyay,pascalutsa}. 
The investigations on which covariant form of the $\Delta$ propagator is more consistent 
with the general principles of quantum field theory are continuing~\cite{kirchbach}.

By using Eqs.(\ref{eq:nr-tpingn}) and (\ref{eq:nr-tpinpin}),
one can write  Eq.(\ref{eq:f-dgn}) as
\begin{eqnarray}
\bar{\Gamma}_{\Delta,\gamma N}= \Gamma_{\Delta,\gamma N}
+ \delta\bar{\Gamma}_{\Delta,\gamma N} \,,
\label{eq:f-dgn-a}
\end{eqnarray}
where
\begin{eqnarray}
\delta\bar{\Gamma}_{\Delta,\gamma N}=
\bar{\Gamma}_{\Delta,\pi N}G_{\pi N}(E)v_{\pi N, \gamma N}  
\label{eq:cloud}
\end{eqnarray}
with
\begin{eqnarray}
\bar{\Gamma}_{\Delta,\pi N}=\Gamma_{\Delta,\pi N}[1 
+ G_{\pi N}(E) t_{\pi N,\pi N}]\,.
\label{eq:f-dpin}
\end{eqnarray}
We will discuss the dynamical content of 
 the dressed vertices Eqs.(\ref{eq:f-dgn-a})-(\ref{eq:f-dpin})
in section III.

Within the considered model the non-resonant $\pi N$ potential
$v_{\pi N,\pi N}$  contains
four  mechanisms: the direct and crossed nucleon terms,
$\rho$ exchange, and crossed $\Delta$ term. Apart from the standard
$\pi NN$ coupling constant, the model is determined by three coupling
constants $g_{\rho NN} g_{\rho\pi\pi}$ and $\kappa_\rho$ for the
$\rho$-exchange, $f_{\pi N\Delta}$ for the $\pi N \rightarrow \Delta$ vertex.
Each interaction vertex is regularized with a
dipole form factor $F(k)=(\Lambda^2/(\Lambda^2+{\vec k}^2))^2$ where
${\vec k}$ is the pion momentum associated with the vertex.
The model thus has additional three parameters: $\Lambda_{\pi NN}$
for the $\pi NN$ vertex, $\Lambda_{\pi N\Delta}$ for the $\pi N \Delta$ vertex and
$\Lambda_\rho$ for both $\rho NN$ and $\rho \pi\pi$ vertices of the 
$\rho$-exchange term. These six parameters are adjusted along with the
bare mass $m_\Delta$ of the $\Delta$ to fit the empirical $\pi N$
scattering phase shifts. We note that this model has considerably
fewer parameters than the other $\pi N$ models used in the
study of pion electroproduction calculations. For example, the $\pi N$ 
model~\cite{hung} used in the Dubna-Mainz-Taipei (DMT) model of pion
electroproduction  has 16 parameters. Thus we do not attempt to fit
the data at energies above $T_L \sim 250$ MeV (invariant mass $W\sim 1280$ MeV).
At higher energies, we expect that the coupling with 
the $\pi\pi N$ channel as well as the tails of higher mass $N^*$, such as
the $P_{11}(1440)$ and $S_{11}(1535)$ resonances,  must be included in
a realistic description of the $\pi N $ scattering data.
An extension of the SL model to higher energies  has been developed recently
in Ref.~\cite{msl}.

\begin{figure}[tb]
\vspace{10pt}
\begin{center}
\mbox{\epsfig{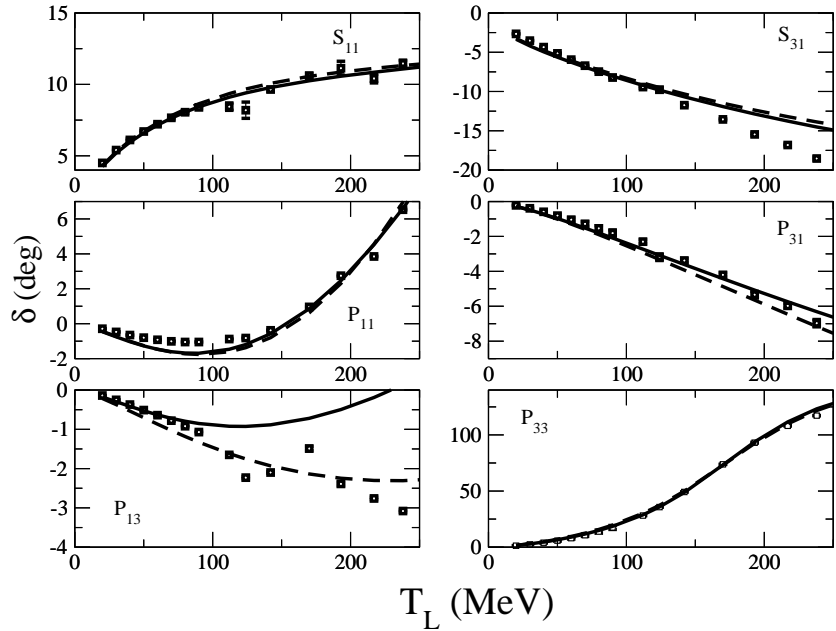}}
\end{center}
\caption{Phase shifts of $\pi N$ elastic scattering up to $T_L=$250 MeV. 
Solid and dotted stand for model SL and SL2 respectively. Data, $L_{2T,2J}$, 
are from the energy independent SAID~\protect\cite{said} analysis plus 8 points from 
their energy dependent solution for the $P_{13}$ and $P_{31}$ partial 
waves at lower energies.
\label{fig:pshift}}
\end{figure}

The original SL model (model L in Ref.~\cite{satolee96}) 
provides a quite good description of the $\pi N$ 
scattering phase-shifts up to $T_L\approx 250$ MeV except in the
$P_{13}$ partial wave.
This can be seen from comparing the data with the solid curves
in Fig.~\ref{fig:pshift}.
This discrepancy raised some questions concerning the stability of 
the results with respect to the $\pi N$ fit. 
In this work, we find that the fit to $P_{13}$ phase shifts can be considerably
improved  while 
retaining a similar quality of fit in the remaining partial waves.  
This is done by weighting the data in $P_{13}$ slightly more heavily in the fit.
The resulting fits are the dotted curves in Fig.~\ref{fig:pshift}.
The fitted parameters for the new $\pi N$ model (SL2) and the original SL model 
in Ref.~\cite{satolee96} are compared in table~\ref{tab:parameters}. 
The remaining discrepancy in the $S_{31}$ channel at energies above $T_L\sim 150$ MeV perhaps 
can be resolved if we add additional mechanisms or use a different form factor 
parameterization. We have explored these possibilities and found that the model 
SL2 is the best we can have within the SL model.

It should be noted here that due to the lack of energy independent solutions 
at lower energies in the $P_{13}$ and $P_{31}$ partial waves, we have 
instead used the SAID~\cite{said} energy dependent 
solutions and assigned those points with $\approx 5\%$ errors in the fits.
A more correct procedure is to fit the original $\pi N$ observables.
This however is not pursued in this work.

\begin{table}[t]
\begin{ruledtabular}
\begin{center}
\begin{tabular}{lll}
           & \multicolumn{2}{c}{Model}\\
 Parameter &                       SL2    &   SL \protect~\cite{satolee96}     \\
\hline
$g_{\rho NN}g_{\rho\pi\pi}$     &37.133   &   38.4329     \\
$k_\rho$                        &2.8496   &   1.825      \\
$\Lambda_{\pi N N}$             &3.6177   &   3.2551     \\
$\Lambda_{\rho}$                &7.4639   &  6.2305     \\
$\Lambda_{\pi N \Delta}$        &3.2132   &   3.29      \\
Bare $m_\Delta$                 &1296.9   &   1299.07     \\
$[f_{\pi N \Delta}\sqrt{25/72}]$&1.1664   &   1.207    \\
                                &         &            \\
$g_{\omega NN}$                  &11.85    &   11.82 \\
$G_M(0)$                        &1.86     &   1.85 \\
$G_E(0)$                        &0.0251   &   0.025 \\
\end{tabular}
\caption{Parameters of the two considered models. The notations are explained 
in the text.  \label{tab:parameters}}
\end{center}
\end{ruledtabular}
\end{table}

\section{The $\gamma^* N\rightarrow \Delta$ Form Factors}

\begin{figure}[tb]
\vspace{10pt}
\begin{center}
\mbox{\epsfig{file=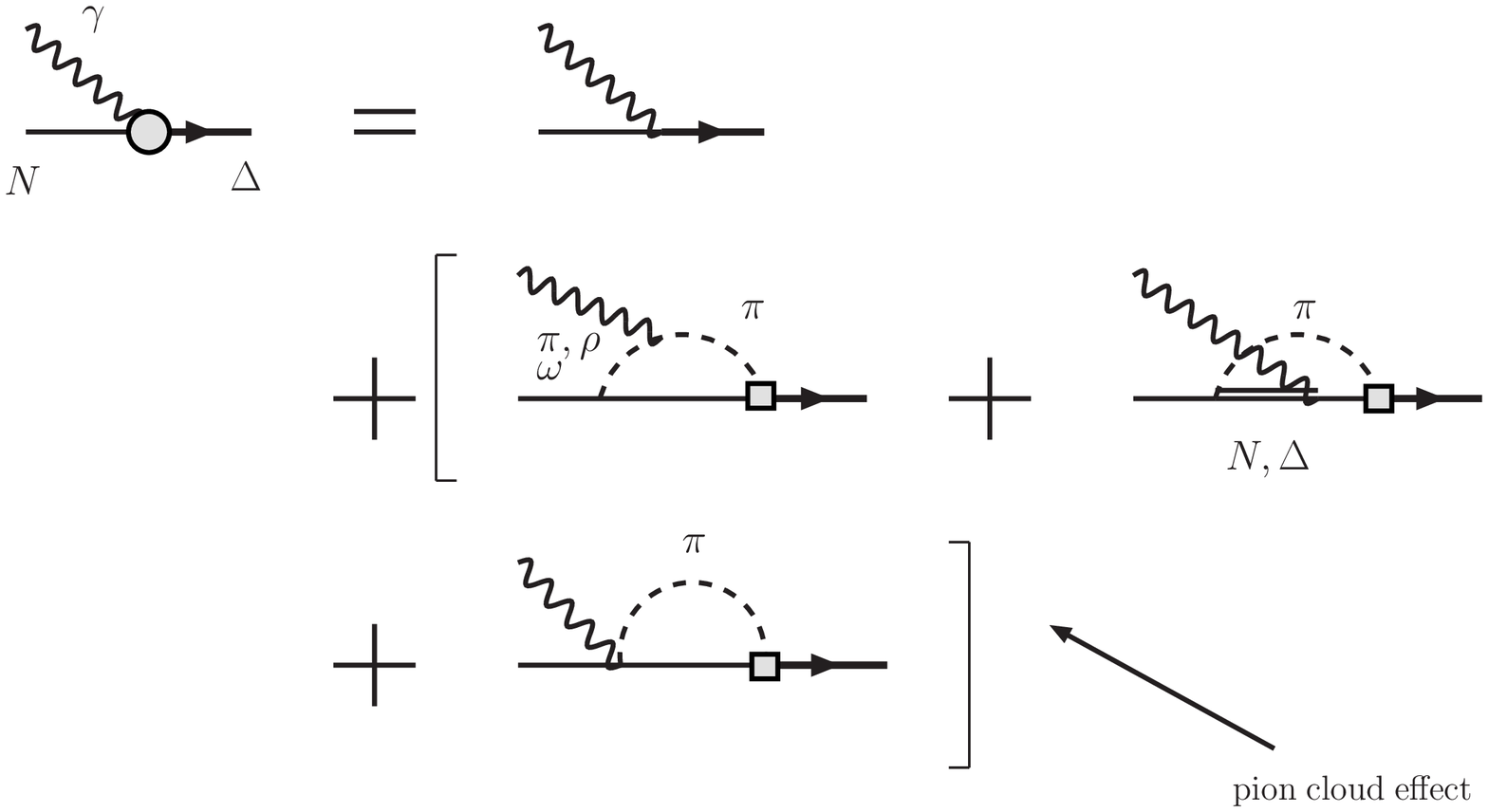, width=15cm}}
\end{center}
\caption{ Graphic representation of the dressed 
$\gamma N \rightarrow \Delta$ vertex defined by Eq.(2.8)-(2.10).
\label{fig:cloud}}
\end{figure}

We now discuss the $\gamma N \rightarrow \Delta$ vertex interaction defined by
Eq.(\ref{eq:f-dgn-a}). It has  a bare vertex 
$\Gamma_{\Delta,\gamma N}$ and 
a term $\delta\bar{\Gamma}_{\Delta,\gamma N}$ which is
determined by the non-resonant interaction $v_{\pi N,\gamma N}$ and a
dressed $\pi N \rightarrow \Delta$ vertex $\bar{\Gamma}_{\Delta,\pi N}$. 
In the SL model, $v_{\pi N,\gamma N}$
contains the usual direct- and crossed-$N$ terms, $\pi$-exchange,
contact term, crossed-$\Delta$ term, 
 and $\rho$- and $\omega$-exchanges.
Thus the dressed $\gamma N \rightarrow \Delta$ contains the meson 
loops illustrated
in Fig.~\ref{fig:cloud}. 
We note that the mechanisms illustrated in  Fig.~\ref{fig:cloud}
are similar to those in the calculations of the current matrix element
$<\Delta|j^{\mu}_{em}\epsilon_\mu | N> $
within a hadron model where $N$ and $\Delta$ contain
a pion cloud: $|B> = Z^{-1/2}[ |B_0> + c_1 |B_0\pi > + \dots]$.
Such calculations can be found, for example, in Ref.~\cite{liu}
using the cloudy bag model.
Thus the term $\delta\bar{\Gamma}_{\Delta,\gamma N}$ is called 
the meson cloud contribution. Note that this definition is not 
shared by the DMT model.

In a consistent derivation using the unitary transformation method
within the SL model,
all strong interaction parameters except the $\omega NN$ vertex in
$v_{\pi N, \gamma N}$ have been determined in  fitting the $\pi N$ scattering 
data.  The $\omega NN$ vertex coupling is determined by using the
pion photoproduction data at $Q^2=0$, as will be further discussed later.
By using the previous works on the $F_{\gamma NN}(Q^2)$, 
$F_{\gamma \pi\pi}(Q^2)$, and $F_{\gamma\pi V}(Q^2)$ 
(with $V=\rho,\omega$) form factors,
the non-resonant interaction $v_{\gamma N,\pi N}$ is then fixed
at any $Q^2$, as explained in Ref.~\cite{satolee01}.
Since the dressed vertex $\bar{\Gamma}_{\Delta,\pi N}$ in Eq.(\ref{eq:cloud})
has also been  fixed in fitting the $\pi N$ scattering data, 
the meson cloud term $\delta\bar{\Gamma}_{\Delta,\gamma N}$
can be predicted at any $Q^2$. With Eq.(\ref{eq:f-dgn-a}),
the bare form factor $\Gamma_{\Delta,\gamma N}$ is then the only unknown,
and can be extracted from the pion electroproduction data.
This strategy of the SL model is also used in this work.
Ideally, one should fit the $\pi N$ and electroproduction data simultaneously
to minimize the errors in extraction. 
But this will be worthwhile only when the polarization
data, such as those  obtained by Kelly et al.~\cite{kel05},
are also included in the fits. Our effort in this direction will be reported
elsewhere.

One of the main objectives in this work is to
explore the theoretical interpretations of the extracted bare 
$\gamma N \rightarrow \Delta$ vertex. We thus need to define this quantity
precisely in the rest of this section.
The bare vertex
$\Gamma_{\Delta,\gamma N}$ of Eq.(\ref{eq:f-dgn}) is parameterized
in the form developed by Jones and Scadron~\cite{js73}. With the
normalization $<{\vec k}|{\vec k'}> = \delta({\vec k}-{\vec k'})$
 for the plane wave 
states and $<\phi_B | \phi_{B'}> = \delta_{B,B'}$ for 
$B = N $ and bare $\Delta$ states, we can write in the rest frame of
the $\Delta$ and for the photon momentum $q=(\omega, {\bf q})$ 
\begin{eqnarray}
 <m_{j_\Delta},  m_{t_\Delta} 
| \Gamma_{\Delta,\gamma N}(q)| \lambda_\gamma\lambda_N, m_{t_N}>
&=& F \langle \thalf m_{t_\Delta} |\ohalf  1 m_{t_N} 0\rangle \nonumber \\
& &\times
[ M_{m_{j_\Delta},\lambda_\gamma\lambda_N}(q) G_M(Q^2) 
+E_{m_{j_\Delta},\lambda_\gamma\lambda_N}(q)G_E(Q^2) \nonumber \\
& & +C_{m_{j_\Delta},\lambda_\gamma\lambda_N}(q)G_C(Q^2)]
\label{eq:mxf}
\end{eqnarray}
where $\lambda_\gamma$ and $\lambda_N$ are the helicities of the
initial photon and nucleon, $m_{j_\Delta}$ is the z-component of
the $\Delta$ spin, $m_{t_\Delta}$ and $m_{t_N}$ denote the isospin
components, and  
\begin{eqnarray}
F=\frac{-e}{(2\pi)^{3/2}}\sqrt{\frac{E_N(\vec{q})+m_N}{2E_N(\vec{ q})}}
\frac{1}{\sqrt{2\omega}}\frac{3(m_\Delta+m_N)}{4m_N(E_N(\vec{q})+m_N)}
\end{eqnarray}
with
\begin{eqnarray}
M_{m_{j_\Delta},\lambda_\gamma\lambda_N}(q) &=&
<m_{j_\Delta}|i\vec{ S}\times \vec{ q}\cdot \vec{ \epsilon}_{\lambda_\gamma}
|\lambda_N> 
\label{eq:coef-m}\\
E_{m_{j_\Delta},\lambda_\gamma\lambda_N}(q) &=&
<m_{j_\Delta}|\vec{ S}\cdot \vec{ \epsilon}_{\lambda_\gamma}
\vec{ \sigma}\cdot \vec{q}
+\vec{ S}\cdot \vec{ q}\vec{ \sigma}\cdot\vec{ \epsilon}_{\lambda_\gamma}|\lambda_N>
 \label{eq:coef-e} \\
C_{m_{j_\Delta},\lambda_\gamma\lambda_N}(q) &=&\frac{1}{m_\Delta}
<m_{j_\Delta}|\vec{S}\cdot\vec{ q}\vec{ \sigma}\cdot\vec{ q}\epsilon_0|
\lambda_N> 
 \label{eq:coef-c}
\end{eqnarray}
where $e=\sqrt{4\pi/137}$, photon polarization vector
is defined by
$\vec{\epsilon}_{\pm 1}=\frac{\mp}{\sqrt{2}}(\hat{x}\pm i \hat{y})$, and
$\epsilon_0 =  0 $ for $\lambda_\gamma =\pm 1$, 
 $\epsilon_0 =  1 $ and $\vec{\epsilon}_0=0$ 
for the scalar component $\lambda_\gamma =0 $.
The transition spin $\vec{S}$ is defined
by $<j_\Delta m_\Delta |S_m|j_Nm_N>=<j_\Delta m_\Delta|j_N 1 m_N m>$.

The form factors $G_M(Q^2)$, $G_E(Q^2)$, and $G_C(Q^2)$ describe magnetic
M1, Electric E2, and Coulomb C2 transitions.  
Choosing the photon direction $\vec{ q}$ in the z-direction,
the above matrix elements
are related to the helicity amplitudes 
defined by Particle Data Group~\cite{pdg} (PDG)
\begin{eqnarray}
A_{3/2}(Q^2) &=& B <m_{j_\Delta}=3/2,  m_{t_\Delta}=m_{t_N}
| \Gamma_{\Delta,\gamma N}(q)| \lambda_\gamma=+1 \lambda_N=-1/2, m_{t_N}> 
\label{eq:he-1}\\
A_{1/2}(Q^2) &=&B<m_{j_\Delta}=1/2,  m_{t_\Delta}=m_{t_N}
| \Gamma_{\Delta,\gamma N}(q)| \lambda_\gamma=+1,\lambda_N=1/2, m_{t_N}>
\label{eq:he2} \\
S_{1/2}(Q^2)&=&B<m_{j_\Delta}=1/2,  m_{t_\Delta}=m_{t_N}
| \Gamma_{\Delta,\gamma N}(q)| \lambda_\gamma=0,\lambda_N=1/2, m_{t_N}>
\label{eq:he-3}
\end{eqnarray}
with
\begin{eqnarray}
B & = & \sqrt{\frac{(2\pi)^3  E_N(\vec{q})\omega}{m_N K_\gamma}}
\end{eqnarray}
where
\begin{eqnarray}
K_\gamma = \frac{m_\Delta^2 - m_N^2}{2m_\Delta}
\end{eqnarray}
With $\vec{ q}$ chosen in the z-direction, we can easily
evaluate Eqs.(\ref{eq:coef-m})-(\ref{eq:coef-c}). Eqs.(\ref{eq:mxf})
and (\ref{eq:he-1})-(\ref{eq:he-3}) then lead to
the following explicit relations
\begin{eqnarray}
A_{3/2}(Q^2) &=& - \frac{\sqrt{3}A}{2} [G_M(Q^2) +  G_E(Q^2)] \,,
\label{eq:he-1a} \\
A_{1/2}(Q^2) &=& - \frac{A}{2}[ G_M(Q^2) - 3 G_E(Q^2)]  \,,
\label{eq:he-2a}\\
S_{1/2}(Q^2)&=& - \frac{|\vec{q}|A}{\sqrt{2}m_\Delta} G_C(Q^2) \,.
\label{eq:he-3a}
\end{eqnarray}
with
\begin{eqnarray}
 A & = & \frac{e}{2m_N}\sqrt{\frac{m_\Delta }{m_N K_\gamma}}
       \frac{|\vec{q}|}{1 + Q^2/(m_N+m_\Delta)^2}
\end{eqnarray}

The helicity amplitudes Eqs.(\ref{eq:he-1})-(\ref{eq:he-3}) are most
often calculated in hadron structure calculations. 
Eqs.(\ref{eq:he-1a})-(\ref{eq:he-3a}) then allow comparisons of
such calculations with 
the bare form factors $G_M(Q^2)$, $G_E(Q^2)$, $G_C(Q^2)$ extracted from our
analyses.
These form factor are
presumably sensitive to the short-range interquark interactions.
They perhaps mainly contain information about
the quark wavefunctions of the $N$ and $\Delta$ within the
constituent quark model, as will be discussed in section V.

In Ref.~\cite{satolee01}, 
it was found that the pion photoproduction data~\cite{legs,mainz} and
the electroproduction data at $Q^2=2.8, 4.0$~GeV$^2$~\cite{fro99} can be fitted well by using the
following naive parameterization
\begin{equation}
G_x(Q^2)=G_x(0)\left(\frac{1}{1+Q^2/0.71({\rm GeV}/c)^2}
\right)^2 (1+a\,Q^2)\,\exp(-b\,Q^2)
\label{eq:par-1}
\end{equation}
where $x=M,E,C$, $a=0.154$ (GeV/c)$^2$ 
and $b=0.166$ (GeV/c)$^2$. The strengths at $Q^2=0$
are found to be $G_M(0)=1.85$, $G_E(0) = 0.025$.  
The value of $G_C(0)$ was fixed using the long wavelength limit
\begin{equation}
G_C(0)=-4 \frac{m^2_{\Delta}}{(m^2_{\Delta}-m^2_N)} G_E(0)
\label{eq:par-2}
\end{equation}

Note that this ${\it ansatz}$ assumes an identical $Q^2$ 
dependence for all three
'bare' electromagnetic couplings. There is no theoretical justification
for this simple choice. Thus it is not clear
whether the discrepancies between the
predictions from the SL model (using the parameterization
Eqs.(\ref{eq:par-1})-(\ref{eq:par-2}) and the data accumulated since 2001
reflect information about
the bare couplings or about deficiencies in 
the model description of the $\pi N$ rescattering process.
With more data in a wide range of $Q^2$, we will abandon in this work the
parameterization Eqs.(\ref{eq:par-1}-\ref{eq:par-2}) and
directly extract these bare couplings
by fitting the data at each individual $Q^2$ point.

The dressed form factor $\bar{\Gamma}_{\Delta,\gamma N}$
has the same symmetry property of the bare vertex defined above. Thus it
can be expanded in the same form of Eq.(\ref{eq:mxf}).
We denote the dressed quantities by $\bar{G}_M(Q^2)$, $\bar{G}_E(Q^2)$,
$\bar{G}_C(Q^2)$. The corresponding helicity amplitudes $\bar{A}_\lambda$
can also be calculated by using the same relations Eqs.(\ref{eq:he-1a})-(\ref{eq:he-3a}).
In Ref.~\cite{satolee01}, it is shown that the dressed ratios can also
be calculated from the imaginary parts of $M_{1+}$, $E_{1+}$ and
$S_{1+}$ amplitudes of pion electroproduction
\begin{eqnarray}
\bar{R}_{EM} &=& - \frac{\bar{G}_E}{\bar{G}_M}= \frac{{\rm Im} E_{1+}}{{\rm Im} M_{1+}} \\
\bar{R}_{SM} &=&  \frac{|\vec{q}|}{2m_\Delta} \frac{\bar{G}_C}{\bar{G}_M}= \frac{{\rm Im} S_{1+}}{{\rm Im} M_{1+}}
\end{eqnarray}
 
It is common to define $G^*_M$ for the M1 transition form factor
which is related to our dressed form factor by
\begin{eqnarray}
G^*_M(Q^2)= \sqrt{\frac{\Gamma^{exp}_\Delta}
{\Gamma^{SL}_\Delta}}\frac{\bar{G}_M(Q^2)}{\sqrt{1+Q^2/(m_\Delta+m_N)^2}}
\label{eq:gmstar}
\end{eqnarray}
where $\Gamma^{exp}_\Delta = 115$ MeV is used in extracting the data from
$M^{3/2}_{1+}$ amplitude of pion electroproduction amplitude and
$\Gamma^{SL}=93 $ MeV from the SL model.

\section{Extraction of $\gamma N \rightarrow \Delta$ Form Factors}

With the refined $\pi N$ model SL2, we first re-analyze the $\pi^0$ 
photoproduction data.
Here we need to also tune the less well determined
parameters of the $\omega$-exchange 
$\gamma N \rightarrow \pi N$ non-resonant interaction. Following the procedure
of Ref.~\cite{satolee96}, we adjust the $\omega NN$ coupling
constant $g_{\omega NN}$ and $G_M(0)$ and $G_E(0)$ of the bare
$N$-$\Delta$ form factor  to fit the data~\cite{legs,mainz} 
of differential cross section 
and photon asymmetry $A_\gamma$ of the $\gamma N \rightarrow \pi N$
reaction. The $\omega NN$ form factor is assumed to be the
same as that of $\rho NN$ which has been fixed in the fit to $\pi N$ data.
The quality of the fit is comparable to what was achieved in
Ref.~\cite{satolee96}, and needs not be discussed here. The 
resulting values of $G_M(0)$ and $G_E(0)$ and $g_{\omega NN}$ are
also listed in the last three lines of Table~\ref{tab:parameters}.

\begin{figure}[t]
%\centerline{\includegraphics[width=10cm]{fig3.eps}}
\mbox{\epsfig{file=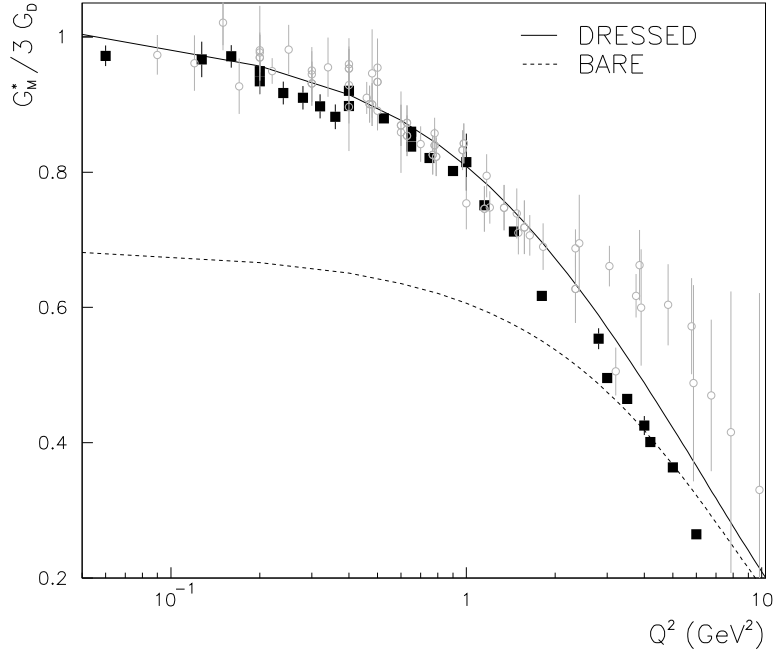, width=11cm}}
\caption{Magnetic dipole transition form factor $G^*_M$
for $\gamma^* N \rightarrow \Delta(1232)$, normalized to
the proton dipole form factor $G_D(Q^2)=1/[1+Q^2/\Lambda^2]^2$ with
$\Lambda^2 = 0.71$ (GeV/c)$^2$.  Experimental points are analyses 
of inclusive data ($\bigcirc$) from pre-1990 experiments at DESY and 
SLAC~\cite{previousgm} and recent exclusive $p(e,e^{\prime}p)\pi^o$ data ($\blacksquare$) from 
BATES~\cite{bates}, MAMI~\cite{Sta06,Spa06} and JLAB~\cite{joo02,smi06,kel05,fro99,ung06}.
Solid curve is from the dressed 
calculation of this work using the parameterization of 
Eqs.~(\ref{eq:par-1})-(\ref{eq:par-2}). The dotted curve
is obtained when the meson cloud effect, defined by Eq.(2.9) is turned off.
\label{fig:gm_Delta}}
\end{figure}

\begin{figure}[t]
\vspace{10pt}
\begin{center}
\mbox{\epsfig{file=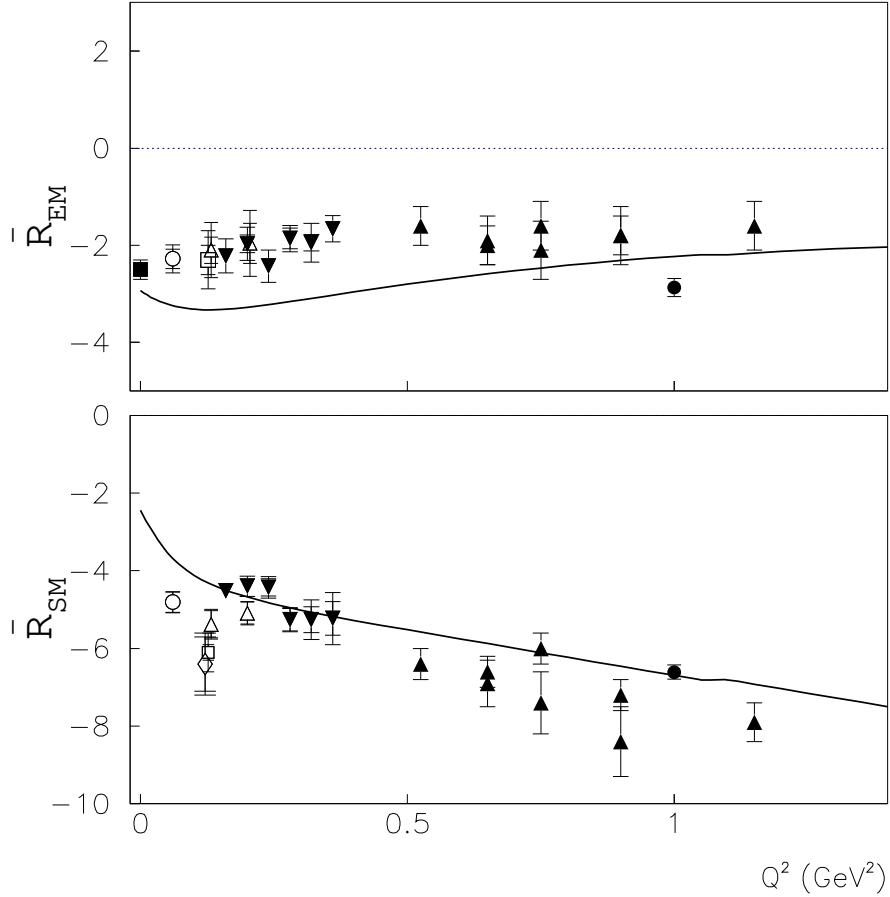, width=13cm}}
\end{center}
\caption{The ratios $\bar{R}_{EM}$ and $\bar{R}_{EM}$ (defined by
Eqs.(3.17)-(3.18)) calculated from the dynamical model using the 
parameterization Eq.(3.15)-(3.16) (solid line).  The data are from recent experiments 
at BATES: $\Box$~\cite{bates}, MAMI: $\blacksquare$~\cite{leu01}, $\bigcirc$~\cite{Sta06}, 
$\bigtriangleup$~\cite{Spa06}, $\Diamond$~\cite{Pos01}, JLAB/CLAS: $\blacktriangledown$~\cite{smi06}, 
$\blacktriangle$~\cite{joo02} and JLAB/Hall A: $\bullet$~\cite{kel05}.  CLAS points 
from Ref.~\cite{smi06} were obtained from the average of UIM and SL model estimates in Table 2.}
\label{fig:mrat_delta}
\end{figure}

Our next step is to extract the $N$-$\Delta$ form factors by fitting
all of the available  pion electroproduction data at energies 
close to the $\Delta$ position.
As a reference point for our analysis, we first use the
parameterization Eqs.(\ref{eq:par-1})-(\ref{eq:par-2}) 
of the SL model to make predictions.
Our results, using that simple parameterization, for the dressed 
M1 form factors $G^*_M(Q^2)$ and the ratios $\bar{R}_{EM}$ and 
$\bar{R}_{SM}$ are the solid curves shown in Figs.~\ref{fig:gm_Delta} 
and~\ref{fig:mrat_delta}. The data shown here are from the analyses performed 
by the experimental groups using mainly the unitary isobar models~\cite{maid,azn03}.
The data from our analyses, shown later in Fig.\ref{fig:lqcd}, are not included in
Figs.~\ref{fig:gm_Delta} and~\ref{fig:mrat_delta}.
The presence of two different extracted values for the same $Q^2$, in these 
and following figures, is due to the existence of two inconsistent data.

It is clear that the resulting dressed $G^*_M(Q^2)$ (solid curve)
 agree well with the available empirical values. In the same figure, we also
show the result (dashed curve) which is obtained by setting
the meson cloud effects, $\delta\bar{\Gamma}_{\Delta,\gamma N}$
defined by Eq.(\ref{eq:cloud}),
to zero. Clearly, the meson cloud effects are important
in the  low $Q^2$ region and gradually diminish as $Q^2$ increases.
The predicted $R_{SM}$ (lower part of Fig.~\ref{fig:mrat_delta})
also agree well with the data except at the two lowest $Q^2$
values $Q^2$ =0.06, 0.127 (GeV/c)$^2$ from the analyses by the MIT-Bates and Mainz 
groups. 

Here we remark on the data presented in 
Figs.~\ref{fig:gm_Delta}-\ref{fig:mrat_delta}.
Because the experiments so far do not have complete measurements including 
all polarization observables, the extraction of multipole amplitudes from the data
needs some constraints imposed by making theoretical assumptions.
The most common practice is to analyze the data starting from
the amplitudes generated from the K-matrix isobar model MAID.
We note that the data from the MIT-Bates and Mainz groups do not have
the full coverage of angles and thus their extracted values of $\bar{R}_{EM}$
and $\bar{R}_{SM}$ perhaps depend very much on the dynamical content in MAID. 
The results from JLab at $Q^2 > 0.16$ (GeV/c)$^2$ are also analyzed using a 
unitarized isobar model (UIM) of Aznauryan~\cite{azn03}.  However the JLAB
data cover almost the whole angular region and hence the fitted results
are closer to a full partial wave analysis.

\begin{figure}[t]
\begin{center}
\mbox{\epsfig{file=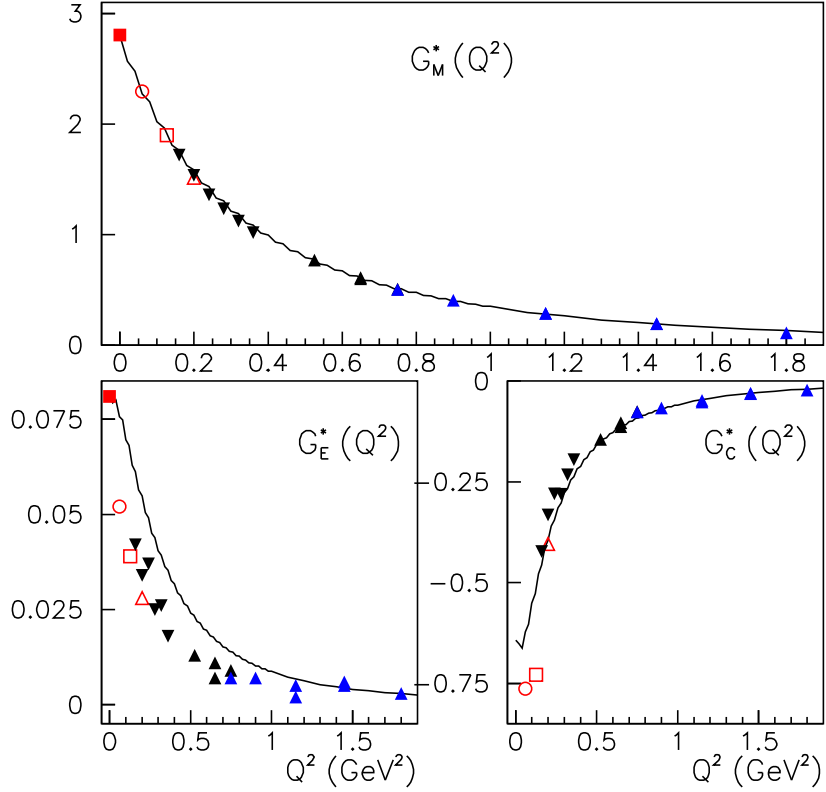, width=120mm}}
\end{center}
\caption{The dressed $\gamma N \rightarrow \Delta$ form factors (solid curves)
calculated using the parameterization 
Eqs.(\ref{eq:par-1})-(\ref{eq:par-2}) are compared with experiment. 
The data are from BATES: $\Box$~\cite{bates}, MAMI: $\blacksquare$~\cite{leu01}, 
$\bigcirc$~\cite{Sta06}, $\bigtriangleup$~\cite{Spa06} and JLAB/CLAS: 
$\blacktriangledown$~\cite{smi06}, $\blacktriangle$~\cite{joo02}.}
\label{fig:dff}
\end{figure}
                                                                                
\begin{figure}[p]
\vspace{10pt}
\begin{center}
\mbox{\epsfig{file=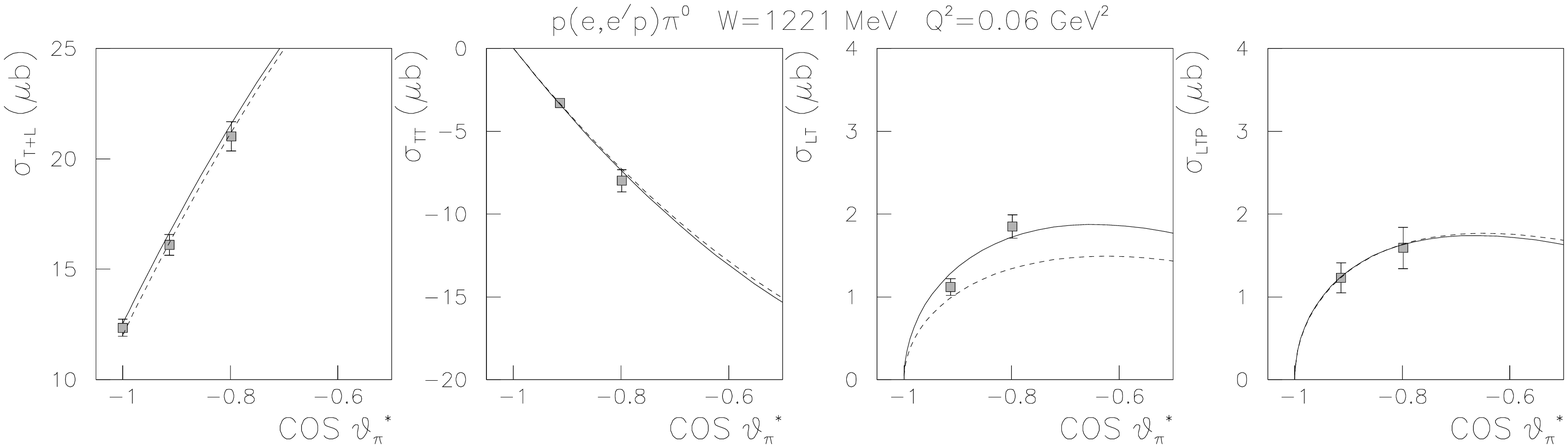, width=12cm}}
\mbox{\epsfig{file=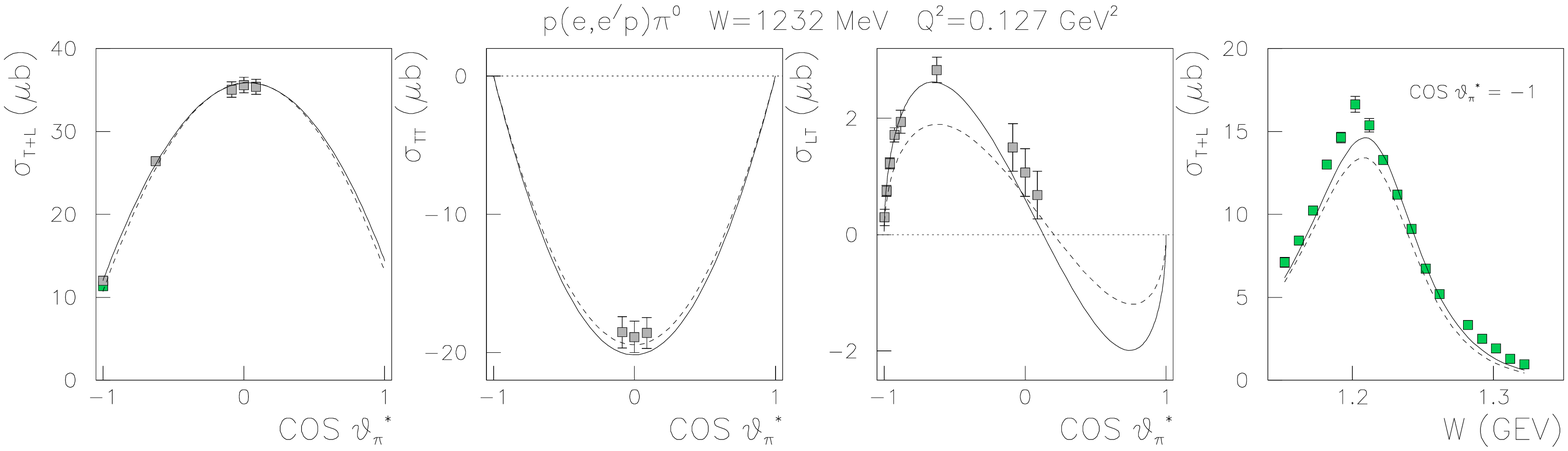, width=12cm}}
\mbox{\epsfig{file=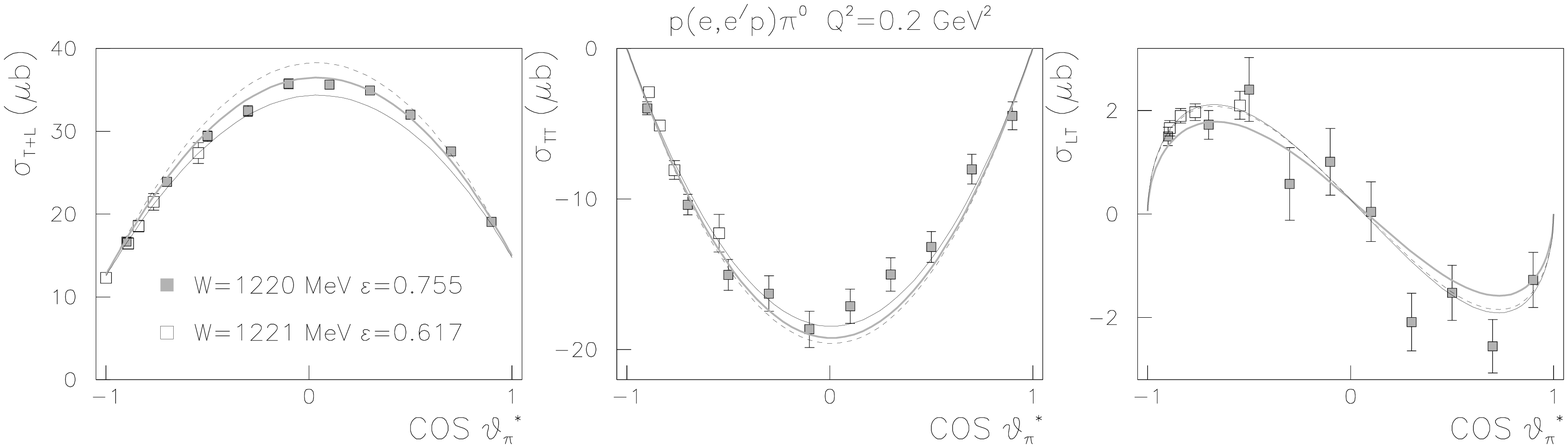, width=12cm}}
\mbox{\epsfig{file=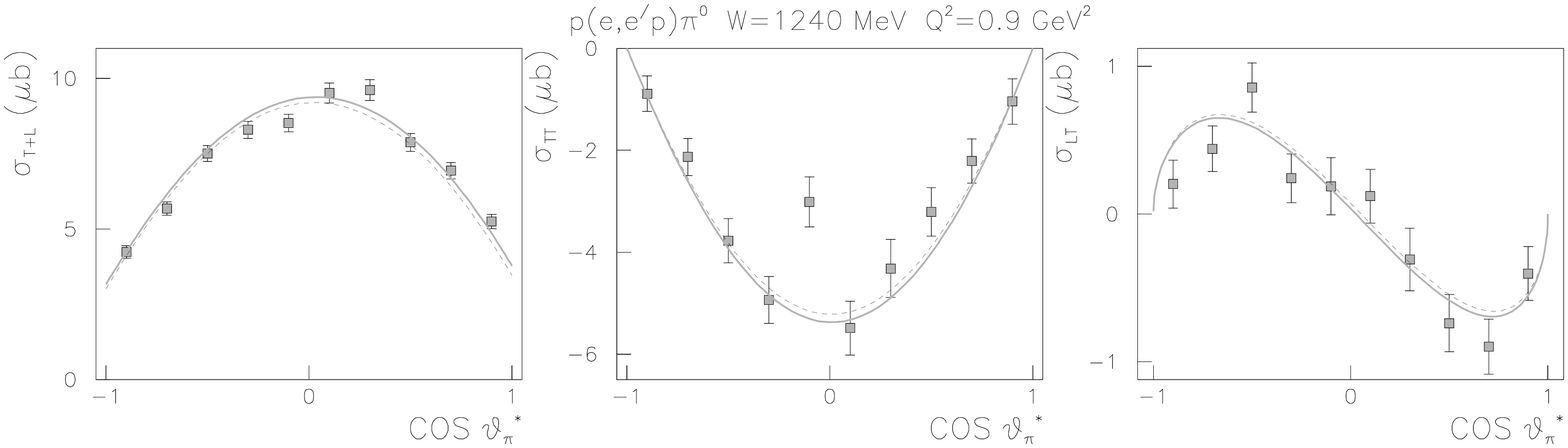, width=12cm}}
\mbox{\epsfig{file=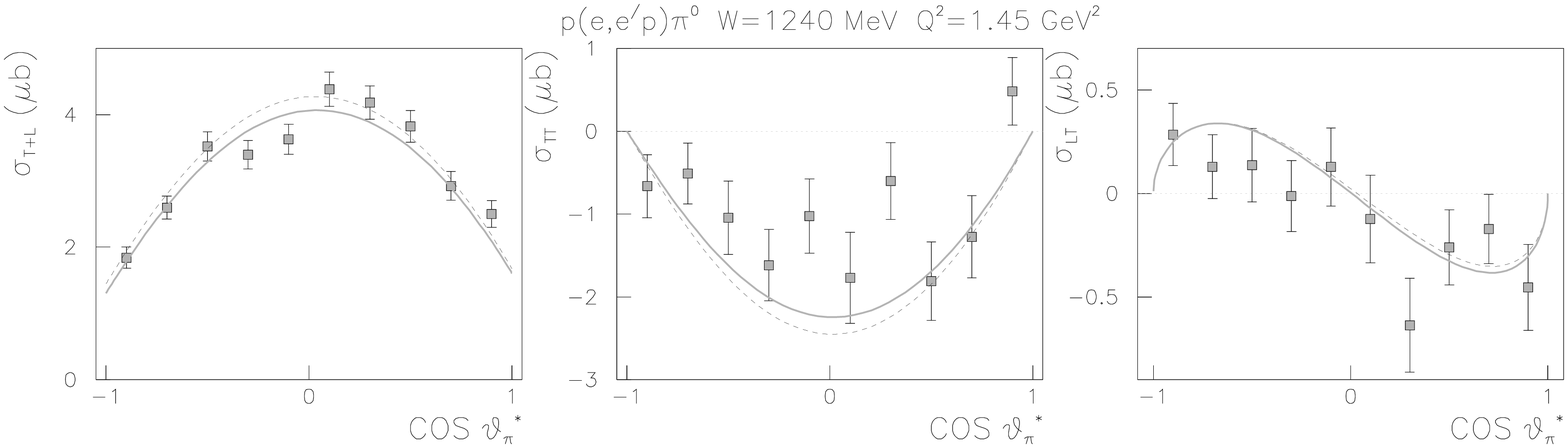, width=12cm}}
\end{center}
\caption{Fits to experimental $p(e,e^{\prime}p)\pi^0$
structure functions. Solid lines are from the fits with
the bare form factors $G_M(Q^2)$, $G_E(Q^2)$ and $G_C(Q^2)$
adjusted at each $Q^2$. The dashed curves are from
the calculations using the parameterization Eqs.(3.15)-(3.16).
Data are from MAMI~\protect\cite{Sta06} at $Q^2=0.06$~GeV$^2$, BATES~\protect\cite{bates}
at $Q^2=0.127$~GeV$^2$, CLAS~\protect\cite{smi06} ($W=1220$~MeV) and MAMI~\protect\cite{Spa06} ($W=1221$~MeV)
at $Q^2=0.2$~GeV$^2$ and CLAS~\protect\cite{joo02} at $Q^2=0.9,1.45$~GeV$^2$.}
\label{fig:slfit1}
\end{figure}

The ratios $\bar{R}_{EM}$ shown in the
upper part of Fig.~\ref{fig:mrat_delta} clearly indicate
that the parameterization Eq.(\ref{eq:par-1})
is not valid for $G_E(Q^2)$. This can be seen more clearly in Fig.~\ref{fig:dff} where
we compare each $N$-$\Delta$ form factor with the empirical values.
To extract more precisely the $N$-$\Delta$ form factors, we therefore
abandon the parameterization Eqs.(\ref{eq:par-1})-(\ref{eq:par-2}) and
perform $\chi^2$ fits to the available differential cross section data
at each $Q^2$ by adjusting the values of the bare form factors.
In Fig.~\ref{fig:slfit1}, we show some results (solid curves) from our fits
at $Q^2= 0.06, 0.127, 0.2, 0.9, 1.45$ (GeV/c)$^2$.
We see that the fits to the differential cross section data
are better than predictions (dotted curves) using the parameterization
Eqs.(\ref{eq:par-1})-(\ref{eq:par-2}). The most visible fit improvement occurs in
the longitudinal-transverse interference cross section $\sigma_{TL}$ 
at $Q^2=0.06, 0.127$ (GeV/c)$^2$

\begin{figure}[h]
\vspace{30pt}
\begin{center}
\mbox{\epsfig{file=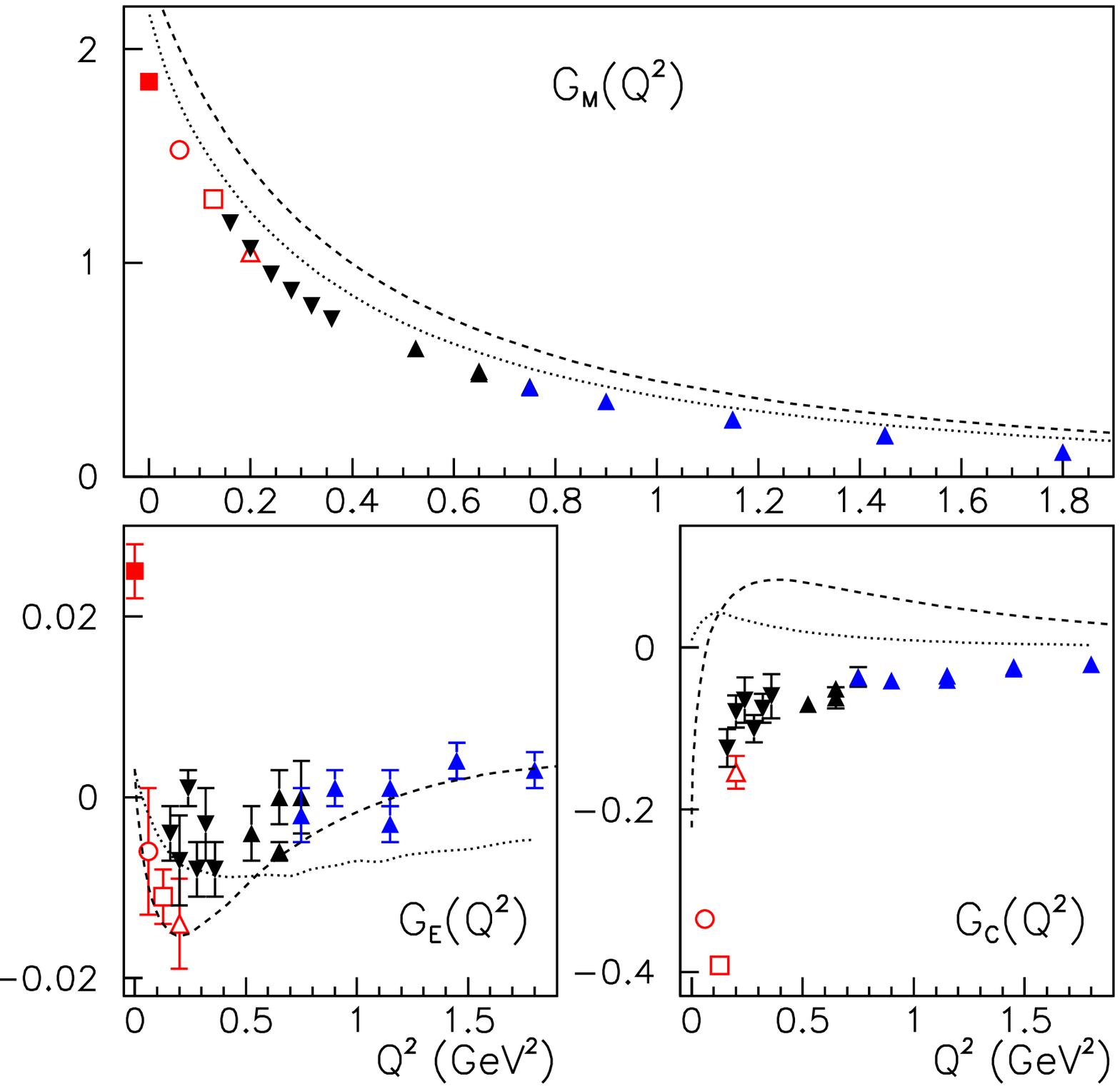, width=110mm}}
\end{center}
\caption{Bare form factors for the $\gamma N \to \Delta$ transition as a function 
of $Q^2$. The points have been obtained by performing individual fits for each $Q^2$ value
to the corresponding pion electroproduction data (BATES: $\Box$~\cite{bates}, MAMI: 
$\blacksquare$~\cite{leu01}, $\bigcirc$~\cite{Sta06}, $\bigtriangleup$~\cite{Spa06} 
and JLAB/CLAS: $\blacktriangledown$~\cite{smi06}, $\blacktriangle$~\cite{joo02}.) 
The dashed curves are from the front form quark model calculations of Refs.~\protect\cite{capstick-1,capstick-c}.
The dotted curves are from the instant form quark model calculations of
Ref.~\protect\cite{bruno-1}.}
\label{fig:bff}
\end{figure}

The resulting bare
form factors are shown in Fig.~\ref{fig:bff}.
We now turn to discussing how these data on the bare 
$\gamma N \rightarrow \Delta$ can be interpreted theoretically.

\section{Theoretical Interpretations}

To explore possible theoretical interpretations of
the extracted $\Delta$ parameters,
it is useful to first consider 
the electromagnetic pion production reaction within
the general reaction theory~\cite{feshbach}. Like any reaction
involving $composite$ systems, such as the atomic and nuclear reactions, its
amplitude
has a non-resonant part and a  resonant part $T = t + t^R$ .
Qualitatively speaking, the non-resonant amplitude $t$ is due to
the {\it fast} process through some direct
particle exchange mechanisms, and the resonant amplitude $t^R$ is
due to the {\it time-delayed} process that the incoming particles lose their
identities and an unstable system is formed and then  decay into various
final states. The unitarity condition ${\rm Im} T = T^\dagger T$ implies that
$t$ and $t^R$ are not independent from each other. In the model considered in this 
work, this can be seen in Eqs.(\ref{eq:rgpi}) for $t^R$ which contain
the non-resonant amplitude $t$ through Eqs.(\ref{eq:f-pind})-(\ref{eq:dself}).
Thus the extracted dressed form factors
$\bar{G}_M(Q^2)$, $\bar{G}_C(Q^2)$, $\bar{G}_C(Q^2)$ of the
resonant amplitude can only be compared with
the hadron structure calculations of current matrix element
$<\Delta | j^\mu_{em}\cdot \epsilon_\mu |N> $
which contain the meson loops illustrated
in Fig.~\ref{fig:cloud}. Furthermore, these mesonic effects are dynamically
identical to those in the non-resonant amplitude 
$t_{\pi N,\gamma N} =(1+ t_{\pi N,\pi N})v_{\pi N, \gamma N} $ 
which is equally important
in describing the data of reaction cross sections. We emphasize here that
this close relation between the resonant and non-resonant amplitude is not
specific to the formulation considered in this work, but is the consequence of a
very general unitarity condition.

Clearly, it is a much more difficult task to understand the extracted
dressed $\bar{G}_M(Q^2)$, $\bar{G}_E(Q^2)$, $\bar{G}_C(Q^2)$ form factors
within QCD.  A simpler approach is to assume that the meson loops included
in our model are the dominant mesonic effects resulting from the
spontaneous breaking of chiral symmetry.  It is then reasonable to compare our bare form
factors ${G}_M(Q^2)$, ${G}_E(Q^2)$, ${G}_C(Q^2)$ with hadron structure
calculations in terms of only the constituent quark degrees of freedom.
It is instructive to first consider the naive
s-wave non-relativistic quark model within which the magnetic M1 transitions
have a well known relation $\mu_{\Delta^+ p}/\mu_p=2\sqrt{2}/3$
where 
\begin{eqnarray}
\frac{e}{2m_p}\mu_p = \langle p, m_{s_N}= \ohalf|\sum_{i}\frac{e_i}{2m_q}\sigma_i(z)|p, m_{s_N}=\ohalf\rangle \\
\frac{e}{2m_p} \mu_{\Delta^+ p} = \langle\Delta^+, m_{s_\Delta}=\ohalf |\sum_{i}\frac{e_i}{2m_q}\sigma_i(z)|p, m_{s_N}=\ohalf\rangle 
\end{eqnarray}
Here $\mu_p$ is the proton magnetic moment with physical
value $\mu^{exp}_p= 1+\kappa_p \sim 2.77$, and
 $m_q$ is the constituent quark mass.
From the above relation and the definition Eq.(\ref{eq:mxf}),
one observe that the magnetic M1
form factor of $\gamma N \rightarrow \Delta$
 at $Q^2=0$ can be directly calculated from the
proton magnetic moment
\begin{eqnarray}
G_{M}(0)&=& [\sqrt{2}G_p(0)]\left[\frac{2(E_N(q)+m_N)}{3(m_\Delta+m_N)}\right]
\sqrt{\frac{2E_N(q)}{E_N(q)+m_N}} = 0.84 \mu_p
\end{eqnarray}
where $q=(m^2_\Delta-m^2_N)/2m_\Delta \sim 260$ MeV/c.
We thus observe that our bare value $G_M(0)=1.85$ can be understood
in terms of constituent quark degrees of freedom only when
we  assume that
the nucleon magnetic moment within the constituent quark model
should be about $\sim 20 \%$ less than the empirical value.
The difference is from the meson loops, similar to that illustrated
in Fig.~\ref{fig:cloud}. Such a mesonic correction is indeed close to what
has been found in the cloudy bag model~\cite{thomas}, and is needed
to explain the discrepancy between the data and the predictions
from a covariant model based on the Dyson-Schwinger Equations~\cite{roberts}.
On the other hand, our extracted bare E2 transition form factor
$G_E(0)$ cannot be understood within the
non-relativistic constituent quark 
model. With the tensor force within the conventional one-gluon-exchange, the 
estimated E2 transition of $\gamma N \rightarrow \Delta$
 is known to be negligibly small compared with the value
calculated from our value $G_E(0) = -0.025$.

We now discuss constituent quark model calculations of 
$\gamma N \rightarrow \Delta$ form factors. Since the $Q^2$ considered
here is not very small,
it is essential to perform calculations relativistically.
We focus on two recent results from Capstick and Keister~\cite{capstick-1}
and Julia-Diaz and Riska~\cite{bruno-1}. Both are within the
framework of relativistic quantum mechanics outlined by Dirac~\cite{dirac}.
There are three possible approaches within this framework: instant form,
front form, and point form. The generators of instant form and front form
can be defined within relativistic quantum field theory and hence are more
closely related to the dynamical reaction model considered in this work.
The calculation of Ref.~\cite{capstick-1} is performed within the front form.
The wavefunctions are expanded up to $N=6$ harmonic oscillator basis states
in a variational calculation using a
relativitized Hamiltonian~\cite{capstick-2} which has
 a one-gluon exchange (OGE) short-distance interaction and Y-shaped
string confinement. The tensor interactions and spin-orbit
interactions are included in both the OGE and confining potentials. All of
the potentials are smeared with some quark size, and momentum dependence of
the potentials is parameterized in order to simulate off-shell effects.
A quark form factor is included in this calculation~\cite{capstick-c}.
We assume that this quark form factor is to account for the finite size of
constituent quarks, not phenomenologically to include the meson loops
as illustrated in Fig.~\ref{fig:cloud}.
This conjecture is of course debatable and should be explored in future.

The calculation of Ref.~\cite{bruno-1} starts with wavefunctions parameterized
as $\phi_0(P) = N(1+\frac{P^2}{4b^2})^{-a} $  with 
$P =\sqrt{2(\vec{k}^{\,\,2}+ \vec{q}^{\,\,2})}$, 
where  $\vec{k}$ and $\vec{q}$ are two intrinsic
momenta of the three-quark system, and $N$ is a normalization factor. 
The parameters $b$, $a$ and the constituent
quark mass $m_q$ are adjusted to fit the nucleon electromagnetic form factors.
With a phenomenological addition of d-state wavefunction, the same parameters
are used to predict the
$\gamma N \rightarrow \Delta$  form factors.
Here we only consider their calculation within the instant form, mainly because
it give the best fits to the data.

The comparisons of the predictions of Refs.~\cite{capstick-2}
and~\cite{bruno-1}  are given in 
Fig.~\ref{fig:bff}. Clearly their results are only in qualitative
agreements with the extracted bare form factors. 
In particular, both calculations fail
to reproduce the
very rapid drop of $G_E(Q^2)$ from the photon point $Q^2=0$
to $Q^2=0.06$. Both calculations predict a sign change of $G_C(Q^2)$
 in disagreement with the extracted values.

\begin{figure}[t!]
\vspace{10pt}
\begin{center}
\mbox{\epsfig{file=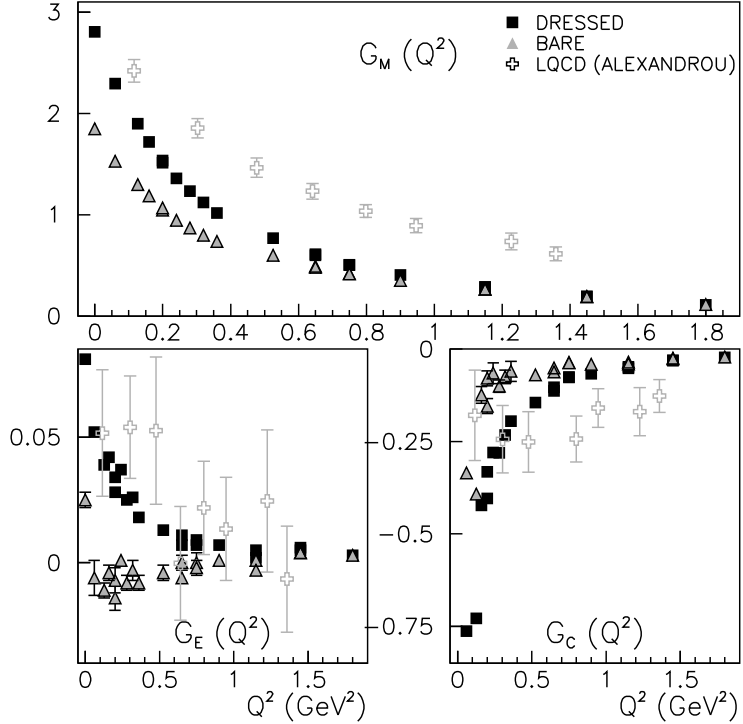, width=110mm}}
\end{center}
\caption{
The extracted $\gamma N \rightarrow \Delta$ form factors.
Dark squares(triangles) are the dressed (bare) values.
Open squares are the lattice QCD calculation of 
Ref.~\protect\cite{alexandrou}.} 
\label{fig:lqcd}
\end{figure}

In the fits (solid curves) shown in Fig.~\ref{fig:slfit1} 
for extracting the bare form factors, 
we of course also extract the dressed form factors. Their
differences, which reflect the meson cloud effects defined
by Eq.(\ref{eq:cloud}),
 can be seen by comparing the solid squares and triangles  in
Fig.~\ref{fig:lqcd}. Obviously, pion cloud effects on
$G_E(Q^2)$ and $G_C(Q^2)$ are very pronounced
at low $Q^2$, as predicted in Ref.~\cite{satolee01}.
In the same figure, we also display the results from
a recent Lattice QCD (LQCD) calculations~\cite{alexandrou}.

We now discuss the results from LQCD calculations.
Currently LQCD can only be performed reliably with very large quark
mass. There are two possible ways to use
the extracted $\gamma N \rightarrow \Delta$ form factors to test
these results.
 The first one is to conjecture that
the meson-loop contributions in the quenched calculations is suppressed
in the calculations with large quark mass and hence their results
should be compared $directly$ with our bare values. 
The second one is to apply the chiral
extrapolation to get results in the physical region with correct current
quark masses. There are two problems in using the chiral extrapolation.
First, it is only valid for low $Q^2$, although it has been used
in a rather high $Q^2$ region. Second, there are higher order corrections
on the commonly used chiral extrapolation, which have not been under
control. 
The uncertainties due to this problem have been discussed by Pascalutsa 
and Vanderhaeghen~\cite{pasc-3}. Thus it is not clear what to conclude 
from Fig.~\ref{fig:lqcd} for the results from LQCD of Ref.~\cite{alexandrou}.
Further investigations are clearly needed.

Finally, we present our determined dressed $\bar{R}_{EM}$ and 
$\bar{R}_{SM}$ in the low $Q^2$ region where very large meson 
cloud effects have been identified in Fig.~\ref{fig:lqcd}. Our 
results are listed in table~\ref{tab:1} and compared with the 
values determined using the unitary isobar model (UIM). The 
difference between our values and that from the UIM reflect some 
model-dependence in the extraction.

\begin{table}
\begin{tabular}{|c|cccc|cccc|}
\hline
  & \multicolumn{4}{|c|}{$\bar{R}_{EM}$(\%)} &  
\multicolumn{4}{|c|}{$\bar{R}_{SM}$(\%)} \\
  $Q^2$ &  & UIM & SL & SL2 &  & UIM & SL & SL2 \\
\hline
0.16 &  & -1.94(0.13) & -2.45(0.2) & -2.57(0.2) & & -4.64(0.19) & -4.44(0.35) & -4.36(0.35) \\
0.20 &  & -1.68(0.18) & -2.21(0.2) & -2.31(0.2) &  & -4.62(0.18) & -4.23(0.35) & -4.14(0.35) \\
0.24 &  & -2.14(0.14) & -2.70(0.2) & -2.76(0.2) &  & -4.60(0.28) & -4.32(0.35) & -4.21(0.35) \\
0.28 &  & -1.69(0.27) & -1.99(0.2) & -2.07(0.2) &  & -5.50(0.31) & -5.08(0.35) & -4.97(0.35) \\
0.32 &  & -1.59(0.17) & -2.29(0.2) & -2.35(0.2) &  & -5.71(0.33) & -4.87(0.35) & -4.75(0.35) \\
0.36 &  & -1.52(0.27) & -1.80(0.2) & -1.82(0.2) &  & -5.79(0.43) & -4.76(0.35) & -4.56(0.35) \\
\hline
\end{tabular}\caption{Extracted values of $E2/M1$ ratio
$\bar{R}_{EM}$ and $C2/M1$ ratio 
$\bar{R}_{SM}=S_{1+}/M_{1+}$ at $Q^2=0.16-0.36$~GeV$^2$
from analysis of preliminary results from a CLAS measurement~\cite{smi06} of the $p(e,e^{\prime}p)\pi^0$ reaction.  Methods used are
 Unitary Isobar Model (UIM) and the SL and SL2 models described in this
work.  Errors are statistical only.}
\label{tab:1}
\end{table}

\section{Discussion and summary}

Within the dynamical model of Refs.~\cite{satolee96,satolee01} 
we have performed a comprehensive
analysis of recent pion electroproduction data to
extract more precisely the $\gamma N \rightarrow \Delta$ form factors.
It is found that the predictions from the original SL model are
not changed much by improving its fit to the $\pi N$ phase shifts in
$P_{13}$ channel. The fits to the very extensive $\pi^0$ electroproduction
data near the $\Delta$ position 
can be improved when the simple parameterization
Eqs.(\ref{eq:par-1})-(\ref{eq:par-2}) is abandoned and the 
strengths of the bare form factors $G_M(Q^2)$, $G_E(Q^2)$, and $G_C(Q^2)$
are adjusted at each $Q^2$ where the data are available. 

In using our results, shown in
Fig.~\ref{fig:lqcd} and table~\ref{tab:1}, to
test hadron structure calculations, it is important to note 
that the differences between the
dressed and bare $\gamma N\rightarrow \Delta$ form factors
are from  the meson loops illustrated in 
Fig.~\ref{fig:cloud}. We emphasize here that the parameters associated
with these loops are the same as that in the non-resonant 
amplitude $t_{\pi N,\gamma N}$ and $t_{\pi N,\pi N}$
which are  important components in
describing both the $\pi N$ scattering phase shifts and pion electroproduction
data. Thus the meson cloud effect identified in this analysis is
well constrained by the {\it reaction} data directly. Such a direct connection
with reaction data is very difficult to achieve in the current
hadron model calculations, in particular the Lattice QCD. 

As an exploratory step, we have
considered two relativistic constituent quark model
calculations of $\gamma N \rightarrow \Delta$ form factors. The
predictions from both models differ significantly from the extracted dressed
values, in particular in the low $Q^2$ region. As seen in Fig.~\ref{fig:bff},
their results for $G_M(Q^2)$ and
$G_E(Q^2)$ qualitatively reproduce our bare values both in shapes and 
magnitudes. This seems to suggest that the hadron calculations without
meson degrees of freedom could make contact with the {\it reaction} data
through the bare from factors extracted from a dynamical analysis such as the
one given in this work. The origin of the
differences in the Coulomb form factor $G_C(Q^2)$
seen in Fig.~\ref{fig:bff} is not clear. 
It will be interesting to explore whether the parameters within
both models can be refined to remove these discrepancies and also improve
their results for the bare $G_M(Q^2)$ and $G_E(Q^2)$. 

It is still difficult to interpret the extracted form factors in terms
 the current LQCD calculations. The discrepancies shown in 
Fig.~\ref{fig:lqcd} indicate the need of a better understanding of the
chiral extrapolation used in obtaining those results as well as 
a significant improvement in LQCD calculations.

To end this paper, we mention that the non-resonant amplitude
$t_{\pi N,\gamma N}$,  which is crucial in  
evaluating the meson cloud effects on the $\gamma N\rightarrow \Delta$ 
form factors,
can be  more sensitively determined in
the energy region away from the $\Delta$ position. 
While the SL model considered in this work can account for the data in the region of
$W\sim 1100-1300$~MeV very well in most cases~\cite{burkertlee04}, 
a more extensive analysis of the data away from the $\Delta$ position
must be performed in the near future when polarization data will
also become more extensive. Our effort in this direction to
further improve our understanding of the $\gamma N \rightarrow \Delta$
form factors is in progress.

\acknowledgements
%\begin{acknowledgments}
B.J-D. wants to thank the hospitality of the theory group at Jefferson 
Laboratory. We also thank Constantia Alexandrou and Simon Capstick 
for sending us their most recent results of $\gamma N \rightarrow \Delta$ 
form factors. This work is supported by the Department of Energy, Office of
Nuclear Physics Division, under contract 
No. DE-AC05-84ER40150 and
Contract No. DE-AC05-060R23177, under which Jefferson Science
Associates operates Jefferson Lab,  Contract No. W-31-109-ENG-38,
the European Hadron Physics Project RII3-CT-2004-506078, 
and by the Japan Society for the Promotion of Science,
Grant-in-Aid for Scientific Research(c) 15540275.
%\end{acknowledgments}

\end{document}